# The effects of copper doping on photocatalytic activity at (101) planes of anatase TiO$_2$: A theoretical study

*M. Hussein N. Assadi[1,*], Dorian A. H. Hanaor[2]*

(1) Graduate School of Engineering Science, Osaka University, Osaka 560−8531, Japan
assadi@aquarius.mp.es.osaka-u.ac.jp, +81668506671
(2) School of Civil Engineering, University of Sydney, NSW 2006, Australia

**Abstract:**

Copper dopants are varyingly reported to enhance photocatalytic activity at titanium dioxide surfaces through uncertain mechanisms. In order to interpret how copper doping might alter the performance of titanium dioxide photocatalysts in aqueous media we applied density functional theory methods to simulate surface units of doped anatase (101) planes. By including van der Waals interactions, we consider the energetics of adsorbed water at anatase surfaces in pristine and copper doped systems. Simulation results indicate that copper dopant at anatase (101) surfaces is most stable in a 2+ oxidation state and a disperse configuration, suggesting the formation of secondary CuO phases is energetically unfavourable. In agreement with previous reports, water at the studied surface is predicted to exhibit molecular adsorption with this tendency slightly enhanced by copper. Results imply that the enhancement of photoactivity at anatase surfaces through Cu doping is more likely to arise from electronic interactions mediated by charge transfer and inter-bandgap states increasing photoexcitation and extending surface-hole lifetimes rather than through the increased density of adsorbed hydroxyl groups.

## 1.    Introduction:

Transition metal doping of titanium dioxide photocatalysts is often carried out with the aim of enhancing photocatalytic activity through improved photoexcitation and/or reduced electron-hole recombination [1,2]. Despite an increasingly large number of investigations into these materials, applied transition metal doped $TiO_2$ photocatalysts are yet to find applications. The entanglement of multiple surface and bulk effects and non-equilibrium behaviour encumber experimental investigations into doped systems and consequently, the mechanisms through which transition metal doping may enhance photocatalytic activity of titanium dioxide remain unclear, limiting progress towards the optimisation and future utilisation of such materials. Moreover, a number of studies have found an impairment to photocatalytic activity may arise from the addition of such dopants, which may act as exciton recombination centres [3,4].

Copper in particular is widely utilised as a dopant in $TiO_2$ in efforts to enhance photocatalytic performance. Relative to pristine materials, Cu doped titanium dioxide (Cu-$TiO_2$) is commonly reported to exhibit improved electron-hole separation with photoexcitation. Although dopant surface concentration would be a more relevant parameter, experimental studies have found improvements in photocatalytic activity to be optimised at Cu dopant levels between 0.1-2 at% [5–11].

Most laboratory studies conducted towards $TiO_2$ photocatalysis have utilised powder form material, predominantly in the anatase phase, although the fabrication of thin films and rutile materials is also often undertaken [12–14]. Copper dopant behaviour in $TiO_2$ is strongly linked to synthesis parameters and the diverse methods employed in materials fabrication and analyses have understandably yielded varied findings. Experimental studies have observed diverse Cu speciation in doped $TiO_2$ with a majority of studies reporting the presence of copper dopant in a $Cu^{2+}$ valence. These species are frequently reported to substitute for $Ti^{4+}$ to give a solution phase with the composition $Cu_xTi_{1-x}O_2$ and an increased lattice density of oxygen vacancies [5,6,10,15–17], and reportedly may also occupy interstices in anatase [11,18]. In addition to solid solutions, $Cu^{2+}$ is widely found to exist in crystalline or amorphous CuO nanoclusters and surface localised $Cu(OH)_2$ in doped $TiO_2$ [5–7,9,19]. Less common than $Cu^{2+}$, the presence of monovalent $Cu^+$ is also reported, both in substitutional positions as well as in $Cu_2O$ clusters [8,11,20]. Reports suggesting the absence of a CuO or $Cu_2O$ phase often rely on XRD data, and it should be noted that such results do not necessarily preclude the existence of amorphous or fine secondary oxide precipitates. Finally, copper dopant as metallic $Cu^0$ species is observed to arise even when fired in air in the absence of a reducing agent and is likely to play a role similar to Au or Ag dopant nanoclusters [8,11].

The experimental evidence for the enhancement of photocatalytic activity in $TiO_2$ through copper doping is inconclusive, and diverse mechanisms are reported. Enhanced activity is most often understood to be driven by reduced charge carrier recombination in Cu-$TiO_2$ materials. This has been reported to arise as the result of photo-generated electrons facilitating the reduction $Cu^{2+} + e^- \rightarrow Cu^+$, thus extending the valence band hole lifetimes at surfaces, which are able to react with adsorbed species to form active radicals [5,6]. However, reports that $Cu^+$ and $Cu^0$ enhance photo-activity more effectively than $Cu^{2+}$ are in disagreement with this mechanism [8,20]. Alternatively, the presence of a CuO or $Cu_2O$ phase in may improve exciton lifetime through electron capture in the secondary phase [21,22] or enhance activity through an increased surface area [5,23]. Metallic clusters are understood to enhance photoactivity by expediting charge transfer with adsorbed species [8] while band-gap narrowing induced by copper cations may extend the range of wavelengths that are utilised in photocatalysis [10,24]. Above a certain threshold, increasing the concentration of Cu dopant is often reported to bring about a diminishment in photocatalytic activity through a combination of enhanced recombination and shading [6,9].

The diversity in the reported behaviour of Cu dopants and their effects on photocatalytic activity stems largely from diverse materials synthesis methods that result in Cu dopant in non-equilibrium sites, as well as from the diverse approaches through which photocatalytic activity is assessed. As with other dopants, the effect of a copper dopant on photoactivity in TiO$_2$ is likely to strongly depend on the experimental conditions used [25]. Additionally, as copper nanostructures are themselves known to act as catalysts for oxidative reactions, decoupling the secondary phase activity from the observed titanium dioxide driven photocatalysis may be problematic from experimental point of view.

In addition to the material interactions of transition metal doping, the speciation of adsorbates at terminated oxide interfaces plays a critical role in controlling the performance of applied photocatalysts and has rarely been considered in computational studies. Photocatalysed phenomena at TiO$_2$ surfaces are most often mediated through adsorbed water and the associated intermediate radical species that arise through photocatalysed redox reactions. Water at TiO$_2$ surfaces in general and anatase surfaces in particular is known to be present though associative (molecular) and dissociative adsorption [26–28]. Considering the likely mode of water adsorption is a key step towards interpreting the performance of doped TiO$_2$ materials as photocatalysts in aqueous media.

Generally speaking, increased water dissociation activity at TiO$_2$ surface gives rise to higher photocatalytic performance in many aqueous target applications, such as water decontamination, owing to the increased presence of photogenerated radicals [29,30]. Favourable dissociation energetics of water adsorbed at the different surfaces of TiO$_2$ are reported to arise as the result of oxygen vacancies, structural defects and the proximity of acidic and basic sites. For this reason, the effect of localised dopants and electronic interactions on the dissociation energetics of surface adsorbed water is of interest towards interpreting dopant effects on photocatalytic performance.

The aim of the present work is to disentangle multiple physical phenomena and shed new light on the effects of copper doping in titanium dioxide photocatalysts by considering multiple aspects of surface interactions from an energetics perspective. This is studied by evaluating the stability of copper at different surface sites in anatase TiO$_2$ including the interaction of surface chemistry and dopant speciation. Through this Density Functional Theory (DFT) based study we assess the mechanisms through which the presence of Cu in TiO$_2$ photocatalysts may give rise to enhanced performance and the circumstances in which this enhancement is anticipated.

2. Methodology

Photocatalysis is a surface mediated phenomenon, and therefore in computational studies towards such applications it is most appropriate to focus on electronic interactions localised at surface regions rather than bulk behaviour, which may differ fundamentally. Furthermore, by focusing on a terminated surface, application-relevant behaviour of titanium dioxide photocatalysts can be computationally studied by including adsorbate interactions in simulations. In the present work we focus our study on the (101) surface plane of anatase phase TiO$_2$ (a-TiO$_2$). The metastable anatase phase is typically the first crystalline phase formed through most synthesis procedures and is the most frequently studied phase in photocatalysis [13,31]. The (101) surface is chosen as this family of crystallographic planes is the most thermodynamically stable and therefore the most appropriate for energetics based computational studies. It also represents the majority of the surface area formed in commonly synthesised materials [32,33].

As illustrated in Fig. 1, the anatase (101) surface was constructed by stacking three a-TiO$_2$ layers with vacuum space of 20 Å added normal to the (101) surface separating consecutive slabs. Water molecules were adsorbed on one side of the slab only while the ions on the outermost layers of the opposite surface were fixed to their bulk positions during the geometry optimisation. In the next step, a 1$a$×4$b$ surface supercell of dimensions 10.90 Å × 15.10 Å was constructed exposing eight outermost O atoms on the surface. Density functional calculations were carried out within augmented plane-wave potential [34,35] formalism as implemented in VASP code [36,37]. Brillouin zone sampling was carried out by choosing a 2×2×1 k-points set within a Monkhorst-Pack scheme that generated a grid with spacing of ~0.05 Å$^{-1}$ between k points while the energy cut-off was set to be 500 eV. Total energy convergence test was performed by increasing the k-point mesh with ~0.03 Å$^{-1}$ spacing; it was found that the total energy differs only by 10$^{-6}$ eV/atom. Additionally, we recalculated the adsorption energy of few configurations of water molecule on the surface in the supercells with stacking of four layers of a-TiO$_2$. We found that the adsorption energy differed by only ~0.1 meV. Therefore, the results were well converged with respect to k-point mesh and supercell thickness. For exchange-correlation functional, we used generalised gradient approximations (GGA) as parameterised by Perdew and Wang [38]. Furthermore, in order to assess the magnitude of the dispersion interaction on the adsorption energies, we also repeated the calculations using the non-local van der Waals (vdW) correlation functional that approximately accounts for dispersion interactions as proposed by Dion [39] and implemented by Klimeš et al. [40] and Roman-Perez et al. [41]. Results of both GGA and vdW functionals are presented in this work.

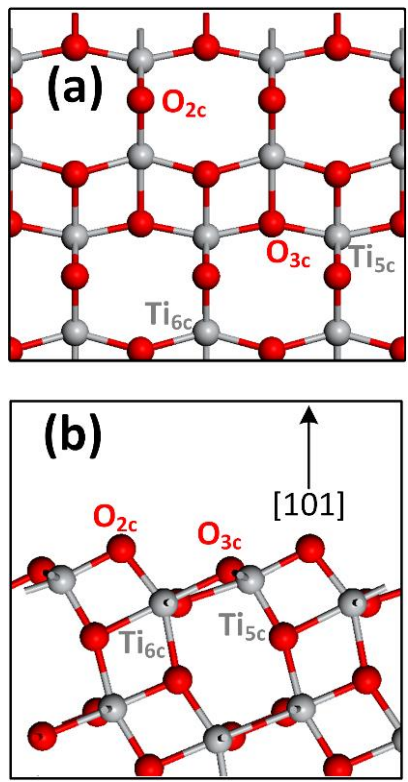

**Fig. 1.** Top (a) and side (b) views of (101) anatase TiO$_2$ are presented. There two distinct sites for each of Ti and O atoms on the outermost layer of the surface. In the case of O, O$_{2c}$ has a coordination number of 2 while O$_{3c}$ has a coordination number of 3. In the case of Ti, Ti$_{5c}$ has a coordination number 5 while Ti$_{6c}$ has a coordination number of 6.

## 3. Results
### 3.1. Water on Pristine a-TiO$_2$ (101)
#### 3.1.1. Associative Water Adsorption

In order to find the most stable configuration of the intact water molecule on the TiO$_2$ anatase (101) surface, we examined all possible configurations in which a water molecule could be positioned at different Ti or O surface sites while simultaneously accounting for all possible orientations of the water molecule with respect to the O-Ti-O bonds of the surface. As presented in Fig. 2, all considered configurations relaxed to three distinct configurations labelled A1−A3. In configuration A1, the water molecule's O atom is located near the Ti$_{5c}$ site. In configuration A2, the water molecule's O atom is still located near the Ti$_{5c}$ site however with a different orientation. In the case of configuration A3, the water molecule's O atom is located near the Ti$_{6c}$ site. In this case the water molecule was only stable in the presented orientation. The water molecule was not found to bind to an O site. The adsorption energy ($E^a$) is defined as follows:

$$E^a = -[E_{conf} - E_{surface} - E_{water}],$$

where $E_{conf}$ is the total energy of the surface and the water molecule for a given configuration, $E_{surface}$ is the total energy of the surface and $E_{water}$ is the total energy of the water molecule. Positive $E^a$ values indicate attractive interaction between the water molecule and the surface. The corresponding adsorption energy of each configuration is given in Table 1. Configuration A1 was found to be the most stable configuration on the TiO$_2$ surface with $E^a$ of 0.87 eV within the GGA framework. In this configuration the distance between water molecule's O atom and the closest Ti atom from the surface ($d_1$) was 2.24 Å. The calculated $E^a$ and $d_1$ of configuration A1 is in good agreement with the previously reported values of 0.82 eV and 2.27 Å for the same configuration [42]. As the distance between the O atom and the closest Ti site ($d_1$) increased in A2 and A3 configurations, the adsorption energy decreased to 0.57 eV and 0.10 eV respectively. When the van der Waals interaction was considered, the adsorption energies increased by a considerable margin; $E^a$ for A1, A2 and A3 configurations were found to be 1.16 eV, 0. 81 eV and 0.23 eV respectively. Nonetheless, the trend of the stability was the same as that of the conventional GGA calculations. The distance between water molecule's O atom the closest Ti atom, however, did not change drastically in the optimised configurations within the framework of vdW interaction.

**Table 1. Adsorption energy (in eV) of associative water molecule adsorption on anatase TiO$_2$ (101) surface.** $d_1$ **is the distance of the O atom in the water molecule to the closest Ti atom on the surface while** $d_2$ **is the shortest distance of water molecules H atom to the closest O atom of the surface.**

| Configuration | GGA | | | vdW | | |
|---|---|---|---|---|---|---|
| | $E^a$ (eV) | $d_1$ (Å) | $d_2$ (Å) | $E^a$ (eV) | $d_1$ (Å) | $d_2$ (Å) |
| A1 | 0.87 | 2.24 | 1.99 | 1.16 | 2.21 | 1.96 |
| A2 | 0.57 | 2.26 | 2.73 | 0.81 | 2.24 | 2.72 |
| A3 | 0.10 | 4.11 | 3.80 | 0.23 | 3.83 | 4.01 |

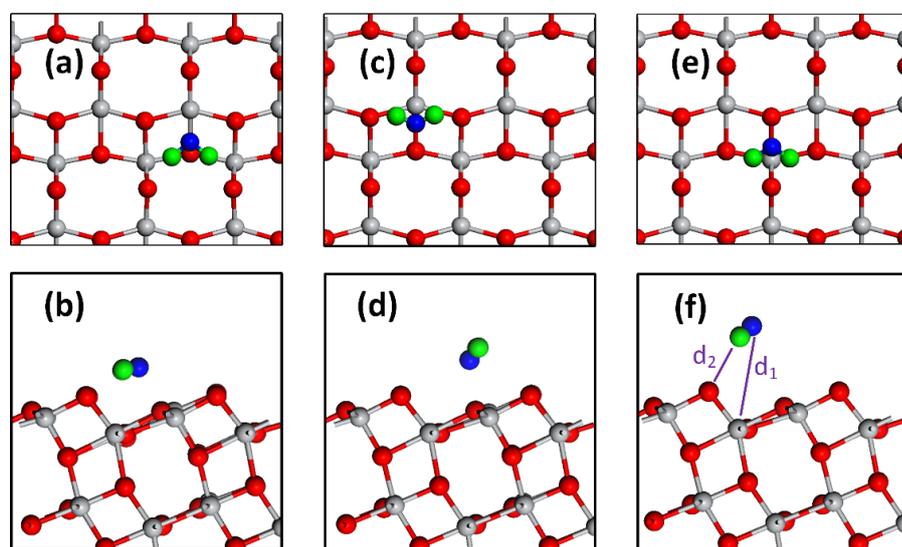

**Fig. 2. Top and side views of the associative adsorption of the water molecule that were found to be stable. (a) and (b) correspond to configurations A1, (c) and (d) correspond to configuration A2 and (e) and (f) correspond to configuration A3.**

### 3.1.2. Dissociative Water Adsorption

In order to study the dissociative water adsorption, we calculated the dissociation energy ($E^d$) of the water molecule based on the most stable associative configuration A1. The dissociation energy is defined as the difference between the total energy of the associative adsorption and the dissociative adsorption. Negative values for $E^d$ indicate that associative adsorption is more stable. The configurations of dissociative adsorption were constructed by splitting the water molecule into OH+H ionic fragments on the a-TiO$_2$ (101) surface. To examine the stability of such dissociation, $E^d$ was calculated for three dissociative configurations (B1−B3) in which the distance between H and O atom in the OH group ($r$) was gradually increased by placing the H atom on the crystallographically distinct O sites at various distances from the OH group. Configurations B1−B3 are presented in Fig. 3 while the dissociation energies are presented in Table 2. Configuration B1 represents the closest possible distance between the OH group and the disassociated H atom with r = 1.95 Å when optimised in GGA scheme.

In this case, as indicated in the isolated surface hydrogen is located on the top of a nearby O$_{3c}$ site with $E^d$ of −0.75 eV. In configuration B2, the isolated hydrogen is located slightly further at a distance of 2.11 Å on the top of an O$_{2c}$ site with an $E^d$ of −0.30 eV. In configuration B3, the isolated H was located at a distance of 4.34 Å on the top of an O$_{3c}$ with an $E^d$ of −0.81 eV. When vdW interaction was included, the dissociation energies all slightly decreased nonetheless still preserving the trend obtained by the GGA. The change in the optimised structures was even more subtle. Since in all cases the $E^d$ was negative, the dissociation of water molecule on the a-TiO$_2$ (101) surface is not thermodynamically favoured. In this regard, undoped (101) surfaces act contrarily to (001) surfaces at which the water molecules are reportedly adsorbed dissociatively [43,44].

**Table 2. Dissociation energy (in eV) of dissociative water molecule adsorption on anatase $TiO_2$ (101) surface.** $d_1$ is the distance between the O molecule in OH group to the closest Ti atom of the surface while $r$ is the distance of the detached H atom to the O atom of the OH group.

| Configuration | GGA | | | vdW | | |
|---|---|---|---|---|---|---|
| | $E^d$ (eV) | $d_1$ (Å) | $r$ (Å) | $E^d$ (eV) | $d_1$ (Å) | $r$ (Å) |
| B1 | −0.75 | 1.87 | 1.89 | −0.82 | 1.86 | 1.95 |
| B2 | −0.30 | 1.87 | 2.11 | −0.31 | 1.87 | 2.15 |
| B3 | −0.81 | 1.81 | 4.34 | −1.05 | 1.81 | 4.42 |

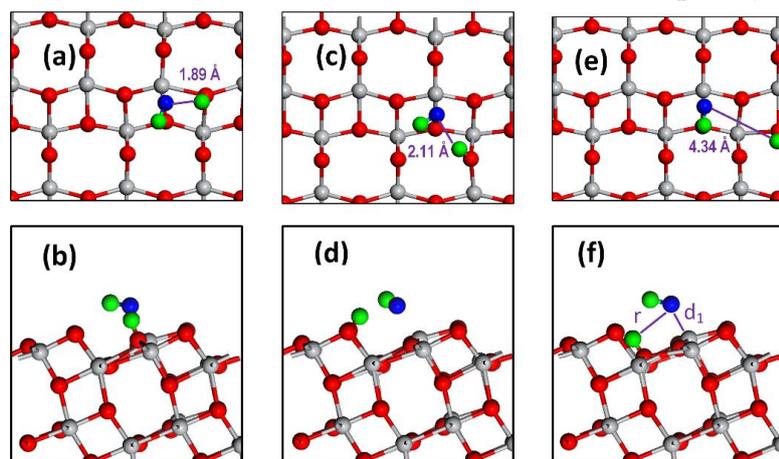

**Fig. 3.** Top and side views of adsorption of sites that were found to be stable for dissociative water molecule. (a) and (b) correspond to configurations B1, (c) and (d) correspond to configuration B2 and (e) and (f) correspond to configuration B3. The distance values presented in the top panels correspond to $r$.

### 3.2. Water adsorption on Cu doped a-$TiO_2$ (101)
#### 3.2.1. Cu on $TiO_2$ Surface

We first identified the most stable substitution site for the Cu atom on the outermost layer of the a-$TiO_2$ (101) surface by comparing the total energies of the configuration in which Cu atoms substitute for $Ti_{5c}$ [Fig. 4(a) and (b)] and the configuration in which Cu substitutes a $Ti_{6c}$ site [Fig. 4(c) and (d)]. Under GGA, Cu substituting for $Ti_{5c}$ was found to be 0.26 eV more stable than the case in which Cu substitutes $Ti_{6c}$. This margin of stability increases to 0.39 eV in the case of vdW inclusive calculations. The density of states (DOS) of both pristine and Cu doped surfaces as calculated by GGA method are presented in Fig. 5(a) and (b) respectively. According to Fig. 5(a), the intrinsic bandgap in $TiO_2$ surface layer is ~2 eV which is considerably smaller than the experimental value of ~3.1 eV. This underestimation is attributed to the GGA's inherent lack of charge derivative discontinuity [45]. Nonetheless, GGA's DOS analysis is still insightful for qualitative evaluation [46]. As demonstrated in Fig. 5(b), the inclusion of Cu on the outermost layer results in an inter-bandgap states in the middle of the fundamental bandgap. This inter-bandgap peak would be expected to manifest optically as a reduced bandgap, and indeed an effective reduction of the bandgap by Cu doping has been previously observed experimentally [11,47].

Furthermore, charge population analysis reveals that there are 8.97 electrons in Cu's d orbitals indicating an oxidation state of 2+ which is in contrast to bulk Cu dopant behaviour [48]. In order to investigate the possibility of Cu aggregation which is a prerequisite for formation of a secondary CuO, we constructed two configurations with two Cu atoms substituting Ti at the surface. In the first configuration the Cu atoms were located at the closest proximity of 3.07 Å from each other while in the second configuration the Cu atoms were located at the furthest distance possible within the supercell of 8.07 Å. We found that the configuration with Cu atoms farthest apart was ~2 meV more stable indicating the absence of a strong tendency towards aggregation among Cu atoms at the a-$TiO_2$ surface.

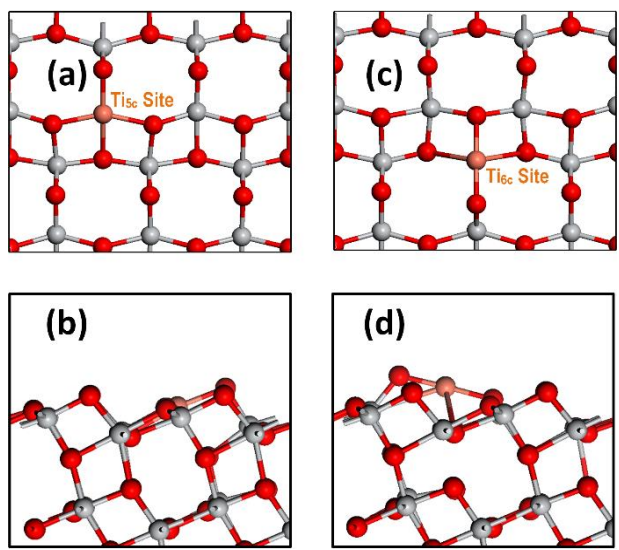

**Fig. 4.** Top and side views of relaxed Cu$_{Ti}$ surface dopant. (a) and (b) correspond to the case where Cu substitutes a Ti$_{5c}$ and (c) and (d) to the case where Cu substitutes aTi$_{6c}$.

### 3.2.2. Associative Water Adsorption

We examined the effect of Cu substitution on the water molecule's associative adsorption by calculating the adsorption energy of a water molecule near a Cu atom that substituted a Ti$_{5c}$. From examining all possible configurations, it was found that, as with pristine surfaces, the A1 configuration is the most stable condition for water adsorption at Cu doped a-TiO$_2$ (101) surface. In this case, $E^a$ was found to be 0.88 eV under GGA and 1.02 eV under vdW calculations. By comparing these values to the $E^a$ values of the A1 configuration in Table 1, we find that the effect of Cu on the adsorption energetics of intact water molecules is insignificant regardless of the computational scheme employed.

### 3.2.3. Dissociative Water Adsorption

The dissociative structure was found to be stable only when the distance of H and the OH group at the beginning of the geometry optimisation was at least 4.90 Å in a similar atomic arrangement to that of configuration B3. When the H atom is located at a closer distance to the OH group in which the initial geometries are similar to those of configurations B1 and B2, due to outward relaxation of the Cu atom in the vicinity of the water molecule, in the optimised geometry, the OH group and the isolated hydrogen atom bind together to form an intact water molecule. The relaxation pattern for the case in which the isolated H atom is located on the top of the O$_{2c}$ site at a distance of 2.75 Å to OH group's O atom is illustrated in Fig. 6. For true dissociative geometry in which the H atom was far enough at 4.90 Å, the dissociation energy was found to be −1.84 eV for the case of GGA and −2.00 eV for the case of vdW functionals. Therefore, the presence of substitutional Cu dopant is found to further inhibit water dissociation at Cu doped a-TiO$_2$ (101) surface.

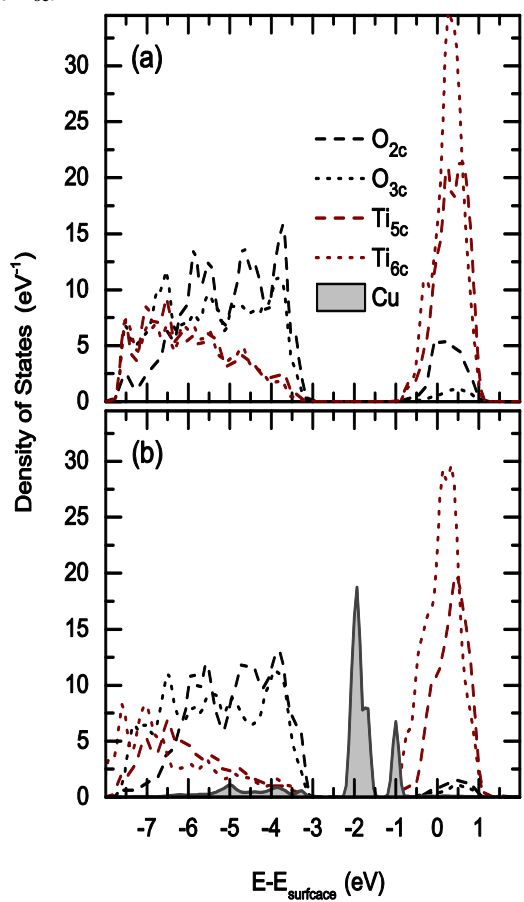

**Fig. 5.** The density of States (DOS) of the surface layer of (a) a-TiO$_2$ and (b) Cu doped a-TiO$_2$ systems with respect to vacuum.

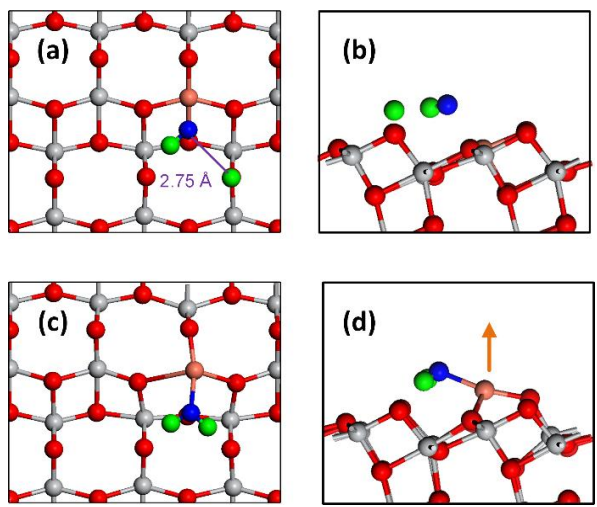

**Fig. 6.** (a) and (b) are the top and side views of the initial and (c) and (d) are the top and side views of the optimised (final) structures of water adsorption on Cu doped a-TiO$_2$ (101) surface. The upward relaxation of Cu atom moves the OH group closer to the isolated H atom that initially was at a distance of 2.75 Å. Consequently, in the optimised geometry the split OH+H complex relaxes into an intact water molecule.

### 3.2.4. Effect of Carrier doping

In order to investigate the effect of photo-induced electron and hole on the adsorption energetics of water, we recalculated the adsorption energies and the dissociation energies in the presence of free carriers. We simulated the presence of a hole in the system by substituting an O atom with a N atom. Likewise, we simulated the presence of an electron in the system by substituting an O atom with a F atom. By using doping as a model for generating carriers, we avoided the large finite size effects that result from the uniform charge background. Furthermore, it has been shown previously that carriers generated by doping accumulate near the surface resembling the photo-generated electrons and holes [49]. To calculate the $E^a$, we used a configuration similar to A1 (most stable configuration for associative water adsorption) and replaced an O atom from the bottom layer with either N or F. Such a site for the carrier dopant eliminates any undesired geometrical effects on the surface. To calculate $E^d$, on the other hand, we constructed a configuration similar to B2 (lowest energy configuration for dissociative water adsorption) with carrier dopants located at the bottom layer. For Cu doped systems, carrier doped structures were based on the most stable associative and dissociative structures. These calculations were performed with GGA functional only and the results are presented in Table 3. In the case of hole doping, the adsorption energy of water molecule on pristine surface remained the same at 0.87 eV while in the case of electron doping, it slightly increased to 0.88 eV. The dissociation energy on the other hand remained the same as that of the undoped system at −0.30 eV. For Cu doped surfaces, the adsorption energy for the hole doped system was 0.81 which is slightly lower than the 0.88 eV of the undoped system while for the electron doped system, the adsorption energy was slightly higher at 0.92 eV. The dissociation energy of the Cu doped surface also varied slightly with carrier doping: In the hole doped system, $E^d$ was −1.77 eV while in the electron doped system $E^d$ was −1.93 eV (for comparison $E^d$ in undoped system for Cu doped surface was −1.84 eV).

**Table 3. The adsorption energy ($E^a$) and the dissociation ($E^d$) energy of water molecule on a-TiO$_2$ and Cu doped a-TiO$_2$ in the present of additional hole and electron.**

| System | N (hole) doping | F (electron) doping |
|---|---|---|
| a-TiO2 (101) surface | $E^a$ = 0.87 eV | $E^a$ = 0.88 eV |
| | $E^d$ = −0.30 eV | $E^d$ = −0.30 eV |
| Cu doped a-TiO2 (101) surface | $E^a$ = 0.81 eV | $E^a$ = 0.92 eV |
| | $E^d$ = −1.77 eV | $E^d$ = −1.93 eV |

### 4. Discussion

The present study focuses on the physico-chemical aspects of anatase surfaces in pristine and copper doped materials. In particular, we focused on interactions with adsorbed water at a surface unit representing the (101) family of planes, the most energetically favourable and widely present crystallographic surface in anatase materials [28]. This methodology is in contrast to most DFT based computational studies which invoke bulk material rather than surface lattice systems.

Results from both GGA and GGA+vdW functionals indicate that at pristine anatase (101) surfaces the presence of adsorbed water is energetically favourable in a molecular or associative type adsorption rather than dissociative, for all adsorption configurations. This trend is consistent with the reported molecular water adsorption from earlier experimental and computational studies [44,50,51]. This may suggest lower levels of hydroxyl radical formation at this crystallographic surface and an increased dominance of pathways mediated through adsorbed molecular species.

The presence of a copper dopant at anatase (101) surfaces is found to be most stable in a disperse configuration suggesting the formation of secondary CuO phases is energetically unfavourable. This does not preclude the formation of secondary phases in the bulk and furthermore synthesis dependant non-equilibrium behaviour is likely to occur owing to inhomogeneous dopant distribution arising during processing or crystallisation. The presence of a substitutional copper dopant, most stable at five-coordinated Ti sites, is not found to result in increased water dissociation activity relative to the pristine material.

Surface dispersed Cu species are predicted to be substitutionally present in a 2+ valence states. This behaviour of Cu in terms of valence and lattice distribution is supported by the experimental findings of Li et al. [6] who concluded that metal to metal charge transfer in Ti-O-Cu complexes at TiO$_2$ surfaces enhance photoactivity through charge carrier separation while CuO clusters serve as exciton recombination centres. Additionally, our simulation results imply that as has been suggested by experimental studies [5,6], the reduction $Cu^{2+} + e^- \rightarrow Cu^+$ may serve as a further mechanism through which the lifetime of surface holes may be extended . This has been supported previously by spin-resonance [52] and diffusive reflectance spectroscopic observations [53]. Increased hole lifetime in the presence of adsorbed water may enhance photoactivity mediated through the reaction $2h^+ + H_2O \rightarrow ½O_2 + 2H^+$.

The introduction of Cu dopants is predicted in the present work to give rise to significant inter-bandgap states, which can be interpreted experimentally as bandgap narrowing. This prediction is in consistency with existing experimental reports and suggests that copper doping may impart visible light activity in anatase photocatalysts, as suggested in previous reports [10,24][47,54]. From simulated carrier doping it was found that neither the presence of a hole nor an electron significantly alters the $E^a$ and $E^d$ values in pristine and copper doped materials suggesting photoexcitation does not impart an increased propensity towards water dissociation at anatase (101) surfaces.

## 5.    Conclusions

We used comprehensive density functional theory based on both GGA and GGA+vdW functionals to study the adsorption and dissociation of water molecules at anatase (101) surfaces, both pristine and doped with a copper dopant. The major findings can be summarised as follows:

a)      We found the most stable site of associative water adsorption on a-TiO2 (101) surface. The adsorption energy of water on this site was 0.87 eV within the GGA functional and 1.16 eV when vdW correction is applied.

b)      Dissociative water adsorption was found to be less energetically favourable having a total energy at least ~0.30 eV higher than the most stable configuration of intact water molecule adsorption.

c)      Cu dopant at a-TiO$_2$ (101) surfaces is most stable when substituting for the Ti$_{5c}$ site. Cu dopants do not notably alter the adsorption energy of associative water molecules and further inhibit water dissociation.

d)      The presence of carriers either holes or electrons does not meaningfully affect the $E^a$ and $E^d$ values of water molecules on pristine or doped surfaces suggesting photoexcitation does not cause further water dissociation.

e)      The presence of substitutional copper dopants result in inter-bandgap states, which impart visible-light photoactivity in anatase.

f)      The calculations with GGA and GGA+vdW functionals render the same trends, although for adsorption energies tend to be higher and dissociation energies tend to be lower with vdW corrected functionals.

Copper doping is often reported to result in enhanced photocatalytic activity in anatase,

the most stable facets of which are the (101) surface. This work suggests that beneficial effects of copper doping in an equilibrium state at anatase (101) surface are more likely to arise through electronic interactions mediated through mechanisms of charge transfer and effective bandgap narrowing, rather than through the modification of surface chemistry in adsorbed water. The existence of secondary copper-bearing phases can not be excluded, as many synthesised materials are not in absolute thermodynamic equilibrium and the role of these phases merits further theoretical investigations. The present work did not consider higher energy planes, such as anatase (001), which often exhibit higher photactivity. The effects of dopants at such surfaces may differ from those found here. However, since the synthesis techniques used to give rise to high levels of (001) planes typically yield non-equilibrium crystallisation and growth behaviour, thorough theoretical and experimental investigation is required to identify the role of Cu dopants on those surfaces.

**Acknowledgements**

**References**

[1] K. Wilke, H.D. Breuer, The influence of transition metal doping on the physical and photocatalytic properties of titania, J. Photochem. Photobio. A 121 (1999) 49−53.

[2] D.F. Zhang, Enhanced photocatalytic activity for titanium dioxide by co-modification with copper and iron, Transit. Metal Chem. 35 (2010) 933−938.

[3] N. Wetchakun, K. Chiang, R. Amal, S. Phanichphant, Synthesis and characterization of transition metal ion doping on the photocatalytic activity of $TiO_2$ nanoparticles, In the proceedings of the IEEE Nanoelectronics Conference, Shanghai, China, 24−27 Mar 2008; IEEE 2008, 43−47.

[4] D.W. Bahnemann, S.N. Kholuiskaya, R. Dillert, A.I. Kulak, A. Kokorin, Photodestruction of dichloroacetic acid catalyzed by nano-sized $TiO_2$ particles, Appl. Catal. B: Environ, 36 (2002) 161−169.

[5] X.H. Xia, Y. Gao, Z. Wang, Z.J. Jia, Structure and photocatalytic properties of copperdoped rutile $TiO_2$ prepared by a low-temperature process, J. Phys. Chem. Solids 69 (2008) 2888−2893.

[6] G. Li, N.M. Dimitrijevic, L. Chen, T. Rajh, K. Gray, A. Role of Surface/Interfacial $Cu^{2+}$ Sites in the Photocatalytic Activity of Coupled CuO−$TiO_2$ Nanocomposites, J. Phys. Chem. C 112 (2008) 19040−19044.

[7] T. Morikawa, Y. Irokawa, T. Ohwaki, Enhanced photocatalytic activity of $TiO_{2-x}N_x$ loaded with copper ions under visible light irradiation, Appl. Catal. A: Gen. 314 (2006) 123−127.

[8] K.Y. Song, Y.T. Kwon, G.J. Choi, W.I. Lee, Photocatalytic activity of Cu/$TiO_2$ with oxidation state of surface-loaded copper, Bull. Korean Chem. Soc. 20 (1999) 957−960.

[9] W. Zhang, Y. Li, S. Zhu, F. Wang, Copper doping in titanium oxide catalyst film prepared by DC reactive magnetron sputtering, Catal. Today 93 (2004) 589−594.

[10] K. Song, J. Zhou, J. Bao, Y. Feng, Photocatalytic Activity of (Copper, Nitrogen)-Codoped Titanium Dioxide Nanoparticles, J. Am. Ceram. Soc. 91 (2008) 1369−1371.

[11] R. López, R. Gómez, M.E. Llanos, Photophysical and photocatalytic properties of nanosized copper-doped titania sol–gel catalysts, Catal. Today 148 (2009) 103−108.

[12] D.A.H. Hanaor, G. Triani, C.C. Sorrell Morphology and photocatalytic activity of highly oriented mixed phase titanium dioxide thin films, Surf. Coat. Technol. 205 (2011) 3658−3664.

[13] D.A.H. Hanaor, C.C. Sorrell, Sand Supported Mixed-Phase $TiO_2$ Photocatalysts for Water Decontamination Applications, Adv. Eng. Mater. 16 (2014) 248−254.

[14] J. Schneider, M. Matsuoka, M. Takeuchi, J. Zhang, Y. Horiuchi, M. Anpo, and D.W. Bahnemann, Understanding $TiO_2$ photocatalysis: mechanisms and materials, Chem. Rev. 114 (2014) 9919−9986.

[15] G. Sivalingam, K. Nagaveni, M.S. Hegde, G. Madras, Photocatalytic degradation of

various dyes by combustion synthesized nano anatase $TiO_2$, App. Catal. B: Environ 45 (2003) 23−38.

[16] J.N. Nian, S.A. Chen, C.C. Tsai, H. Teng, Structural feature and catalytic performance of Cu species distributed over $TiO_2$ nanotubes, J. Chem. Phys. B, 110 (2006) 25817−25824.

[17] D.A.H. Hanaor, C.C. Sorrell, Review of the anatase to rutile phase transformation, J. Mater. Sci. 46 (2011) 855−874.

[18] V.S. Grunin, G.D. Davtyan, V.A. Ioffe, I.B. Patrina, EPR of $Cu^{2+}$ and radiation centres in anatase ($TiO_2$), Phys. Status Solidi B 77 (1976) 85−92.

[19] M.S.P. Francisco, P.A.P. Nascente, V.R. Mastelaro, A.O. Florentino, X-ray photoelectron spectroscopy, x-ray absorption spectroscopy, and x-ray diffraction characterization of $CuO–TiO_2–CeO_2$ catalyst system, J. Vac. Sci. Technol. A 19 (2001) 1150−1157.

[20] G. Colón, M. Maicu, M.C. Hidalgo, J.A. Navío, Cu-doped $TiO_2$ systems with improved photocatalytic activity, Appl. Catal. B 67 (2006) 41−51.

[21] M. Liu, X. Qiu, M. Miyauchi, K. Hashimoto, Cu(II) oxide amorphous nanoclusters grafted $Ti^{3+}$ self-doped $TiO_2$: an efficient visible light photocatalyst, Chem. Mater. 23 (2011) 5282−5286.

[22] Q. Hu, J. Huang, G.Li, b, J. Chen, Z. Zhang, Z. Deng, Y. Jiang, W. Guo, Y. Cao, Effective water splitting using $CuO_x/TiO_2$ composite films: Role of Cu species and content in hydrogen generation, Appl. Surf. Sci. 369 (2016) 201−206.

[23] M.S.P. Francisco, V.R. Mastelaro, P.A.P. Nascente, A.O. Florentino, Activity and characterization by XPS, HR-TEM, Raman spectroscopy, and BET surface area of $CuO/CeO2-TiO_2$ catalysts, J. Phys. Chem. B 105 (2001) 10515−10522.

[24] B. Choudhury, M. Dey, A. Choudhury, Defect generation, d-d transition, and band gap reduction in Cu-doped $TiO_2$ nanoparticles, Int. Nano Lett. 3 (2013) 1−8.

[25] V. Subramanian, E.E. Wolf, P.V. Kamat, Influence of metal/metal ion concentration on the photocatalytic activity of $TiO_2$-Au composite nanoparticles, Langmuir, 19 (2003) 469−474.

[26] A.V. Bandura, D.G. Sykes, V. Shapovalov, T.N. Troung, J.D. Kubicki, R.A. Evarestov, Adsorption of Water on the $TiO_2$ (Rutile) (110) Surface: A Comparison of Periodic and Embedded Cluster Calculations, J. Phys. Chem. B 108 (2004) 7844−7853.

[27] L.E. Walle, A. Borg, E.M. Johansson, S. Plogmaker, H. Rensmo, P. Uvdal, A. Sandell, Mixed Dissociative and Molecular Water Adsorption on Anatase $TiO_2$(101), J. Phys. Chem. C 115 (2011) 9545−9550.

[28] H. Cheng, A. Selloni, Hydroxide Ions at the Water/Anatase $TiO_2$(101) Interface: Structure and Electronic States from First Principles Molecular Dynamics, Langmuir 26 (2010) 11518−11525.

[29] M.A. Henderson, Structural Sensitivity in the Dissociation of Water on $TiO_2$ Single-Crystal Surfaces, Langmuir 12 (1996) 5093−5098.

[30] R. Schaub, P. Thostrup, N. Lopez, E. Lægsgaard, I. Stensgaard, J. K. Nørskov, F. Besenbacher, Oxygen Vacancies as Active Sites for Water Dissociation on Rutile $TiO_2$(110). Phys. Rev. Lett. 87 (2001) 266104.

[31] R.D. Shannon, J.A. Pask, Kinetics of the Anatase-Rutile Transformation, J. Am. Ceram. Soc. 48 (1965) 391−398.

[32] X.Q. Gong, A. Selloni, M. Batzill, U. Diebold, Steps on anatase $TiO_2$(101), Nat. Mater. 5 (2006) 665−670.

[33] W. Hebenstreit, N. Ruzycki, G.S. Herman, Y. Gao, U. Diebold, Scanning tunneling microscopy investigation of the $TiO_2$ anatase (101) surface, Phys. Rev. B 62 (2000) R16334.

[34] P.E. Blochl, Projector augmented-wave method, Phys. Rev. B 50 (1994) 17953−17979.

[35] G. Kresse, D. Joubert, From ultrasoft pseudopotentials to the projector augmented-wave method, Phys. Rev. B 59 (1999) 1758−1775.

[36] G. Kresse, J. Furthmuller, Efficient iterative schemes for ab initio total-energy

calculations using a plane-wave basis set, Phys. Rev. B 54 (1996) 11169−11186.

[37] G. Kresse, J. Furthmuller, Efficiency of ab-initio total energy calculations for metals and semiconductors using a plane-wave basis set, Comp. Mater. Sci. 6 (1996) 15−50.

[38] J.P. Perdew, K. Burke, M. Ernzerhof, Generalized Gradient Approximation Made Simple, Phys. Rev. Lett. 77 (1996) 3865−3868.

[39] M. Dion, H. Rydberg, E. Schroder, D.C. Langreth, B.I. Lundqvist, Van der Waals Density Functional for General Geometries, Phys. Rev. Lett. 92 (2004) 246401.

[40] J. Klimes, D.R. Bowler, A. Michaelides, Van der Waals density functionals applied to solids, Phys. Rev. B 83 (2011) 195131.

[41] G. Roman-Perez, J.M. Soler, Efficient Implementation of a van der Waals Density Functional: Application to Double-Wall Carbon Nanotubes, Phys. Rev. Lett. 103 (2009) 096102.

[42] U.J. Aschauer, A. Tilocca, A. Selloni, Ab initio simulations of the structure of thin water layers on defective anatase $TiO_2$(101) surfaces, Int. J. Quantum Chem. 115 (2015) 1250−1257.

[43] X.Q. Gong, A. Selloni, Reactivity of Anatase Nanoparticles: The role of the minority (001) surface, J. Phys. Chem. B 109 (2005) 19560−19562.

[44] A. Vittadini, A. Selloni, F.P. Rotzinger, M. Gratzel, Strucutre and energetics of water adsorbed at $TiO_2$ anatase (101) and (001) surfaces, Phys. Rev. Lett. 81 (1998) 2954−2957.

[45] L. J. Sham M. Schlüter, Density-Functional Theory of the Energy Gap. Phys. Rev. Lett. 51 (1983) 1888−1981.

[46] N. Mardirossian, Head-Gordon, M. Exploring the limit of accuracy for density functionals based on the generalized gradient approximation: Local, global hybrid, and range-separated hybrid functionals with and without dispersion corrections, J. Appl. Phys. 40 (2014) 18A527.

[47] L. Yoong, F.K. Chong, B.K. Dutta, Development of copper-doped $TiO_2$ photocatalyst for hydrogen production under visible light. Energy 34 (2009) 1652−1661.

[48] M.H.N. Assadi, D.A.H. Hanaor, Theoretical study on copper's energetics and magnetism in $TiO_2$ polymorphs, J. Appl. Phys. 113 (2013) 233913.

[49] A. Boonchun, N. Umezawa, T. Ohno, S. Ouyang, J. Ye, Role of photoexcited electrons in hydrogen evolution from platinum co-catalysts loaded on anatase $TiO_2$: a first-principles study. J. Mater. Chem. A 1 (2013) 6664−6669.

[50] G.S. Herman, Z. Dohnalek, N. Ruzycki, U. Diebold, Experimental Investigation of the Interaction of Water and Methanol with Anatase−$TiO_2$(101), J. Phys. Chem. B 107 (2003) 2788−2795.

[51] Y. He, A. Tilocca, O. Dulub, A. Selloni, U. Diebold, Local ordering and electronic signatures of submonolayer water on anatase $TiO_2$(101), Nat. Mater. 8 (2009) 585−589.

[52] C. Chen, X. Li, W. Ma, J. Zhao, H. Hidaka, N. Serpone, Effect of Transition Metal Ions on the $TiO_2$-Assisted Photodegradation of Dyes under Visible Irradiation: A Probe for the Interfacial Electron Transfer Process and Reaction Mechanism, J. Phys. Chem. B 106 (2002) 318−324.

[53] M. Hamadanian, A. Reisi-Vanani, A. Majedi, Synthesis, characterization and effect of calcination temperature on phase transformation and photocatalytic activity of Cu, S-codoped $TiO_2$ nanoparticles, Appl. Surf. Sci. 256 (2010) 1837–1844.

[54] J. Choi, H. Park, M. R. Hoffmann, Effects of Single Metal-Ion Doping on the Visible-Light Photoreactivity of $TiO_2$, J. Phys. Chem. C 114 (2009) 783−792.